\documentclass[preprint,10pt,aps,prl,twocolumn,preprintnumbers,nofootinbib]{revtex4-2}
\usepackage{amsmath,amssymb,mathtools}
\usepackage{hyperref}
\usepackage{xspace}
\usepackage{graphicx}
\usepackage{grffile}
\usepackage{enumitem}
\usepackage{xcolor}
\newcommand{\sect}[1]{%
\bigskip\noindent%
{\bfseries\upshape\rmfamily\boldmath{#1}.}---%
\ignorespaces%
}

\DeclareRobustCommand{\ensuremathrm}[1]{\ensuremath{\mathrm{#1}}}

\DeclareRobustCommand{\NLO}{\text{NLO}\xspace}
\DeclareRobustCommand{\NLOPS}{\text{NLO+PS}\xspace}
\DeclareRobustCommand{\NNLO}{\text{NNLO}\xspace}

\DeclareRobustCommand{\re}{\ensuremathrm{e}}
\DeclareRobustCommand{\rs}{\ensuremathrm{s}}
\DeclareRobustCommand{\rT}{\ensuremathrm{T}}

\DeclareRobustCommand{\jet}{\ensuremathrm{jet}\xspace}
\DeclareRobustCommand{\jets}{\ensuremathrm{jets}\xspace}
\DeclareRobustCommand{\tag}{\ensuremathrm{tag}\xspace}

\DeclareRobustCommand{\alphas}{\ensuremath{\alpha_\rs}\xspace}
\DeclareRobustCommand{\Pp}{{\ensuremathrm{p}}\xspace}

\DeclareRobustCommand{\Pep}{{\ensuremathrm{e^+}}\xspace}
\DeclareRobustCommand{\Pem}{{\ensuremathrm{e^-}}\xspace}

\DeclareRobustCommand{\PD}{{\ensuremathrm{D}}\xspace}
\DeclareRobustCommand{\PB}{{\ensuremathrm{B}}\xspace}
\DeclareRobustCommand{\PZ}{{\ensuremathrm{Z}}\xspace}

\DeclareRobustCommand{\Pg}{{\ensuremathrm{g}}\xspace}
\DeclareRobustCommand{\Pq}{{\ensuremathrm{q}}\xspace}

\DeclareRobustCommand{\Pqb}{{\ensuremathrm{b}}\xspace}

\DeclareRobustCommand{\Pqc}{{\ensuremathrm{c}}\xspace}

\DeclareRobustCommand{\GeV}{\ensuremathrm{GeV}\xspace}
\DeclareRobustCommand{\TeV}{\ensuremathrm{TeV}\xspace}

\DeclareRobustCommand{\Fl}{\ensuremath{\mathfrak{f}}\xspace}
\DeclareRobustCommand{\fl}{\ensuremathrm{f}\xspace}
\DeclareRobustCommand{\afl}{\ensuremath{\bar{\fl}}\xspace}

\DeclareRobustCommand{\dset}{\ensuremath{\mathcal{D}}\xspace}

\DeclareRobustCommand{\aMCatNLO}{\textsc{\small MadGraph5\_aMC@NLO}\xspace}
\DeclareRobustCommand{\pythia}{\text{Pythia}\xspace}

\begin{document}

\preprint{BONN-TH-2022-17, CERN-TH-2022-121, ZU-TH 31/22}

\title{A dress of flavour to suit any jet}%

\author{%
  Rhorry Gauld${}^1$, 
  Alexander Huss${}^2$, 
  Giovanni Stagnitto${}^3$
\bigskip}
\affiliation{%
${}^1$Bethe Center for Theoretical Physics \& Physikalisches Institut der Universit\"at Bonn, D-53115 Bonn, Germany
\\
${}^2$Theoretical Physics Department, CERN, CH-1211 Geneva 23, Switzerland
\\
${}^3$Physik-Institut, Universit\"at Z\"urich, Winterthurerstrasse 190, 8057 Z\"urich, Switzerland
\bigskip}%

\date{\today}

\begin{abstract}
  Identifying the flavour of reconstructed hadronic jets is critical for precision phenomenology and the search for new physics at collider experiments, as it allows to pinpoint specific scattering processes and reject backgrounds. 
  Jet measurements at the LHC are almost universally performed using the anti-$k_\rT$ algorithm, however no approach exists to define the jet flavour for this algorithm that is infrared and collinear (IRC) safe. 
  We propose a new approach, a {\em flavour dressing} algorithm, that is IRC safe in perturbation theory and can be combined with any definition of a jet.
  We test the algorithm in a $\Pep\Pem$ environment, and consider the $\Pp\Pp \to \PZ+\Pqb\text{-}\jet$ process as a practical application.
\end{abstract}

\maketitle

\sect{Introduction}
The confining property of the strong interactions---described by quantum chromodynamics (QCD)---prohibits the observation of free quarks and gluons: in high-energy particle collisions, such as those at the Large Hadron Collider (LHC), they give rise to collimated sprays of hadrons inside the detector, denoted as \emph{jets}.
A jet is defined by the associated reconstruction algorithm and plays a crucial role as the interface between experiment and theory.
In this regard, a core property of any jet algorithm is infrared and collinear (IRC) safety, i.e.\ the insensitivity to soft (low energy) emissions and collinear (small angle) splittings.
Only if such a property is satisfied, can a comparison between measurements and theoretical predictions based upon fixed-order perturbation theory be reliably carried out.

Further identifying the ``flavour'' of the jets is critical to pinpoint specific scattering processes and reject backgrounds.
An important example is the identification of a jet which is consistent with being initiated by a heavy-flavour (charm or beauty) quark. 
The identification of such signatures provides a window into the interactions of heavy-flavour quarks with other fundamental particles from \GeV to \TeV energy scales. 
This in turn provides a unique opportunity to: perform (flavour-specific) direct searches for new physics phenomena~\cite{CMS:2019zmd,ATLAS:2021yij}; test the mechanism for generating the mass of elementary particles~\cite{ATLAS:2018kot,CMS:2018nsn,ATLAS:2019yhn,ATLAS:2020fcp,ATLAS:2020jwz}; probe the internal flavour structure of hadrons~\cite{ATLAS:2014jkm,LHCb:2021stx,CMS:2022bjk}.

Jet measurements at the LHC are almost universally performed using the anti-$k_\rT$ algorithm~\cite{Cacciari:2008gp} owing to the geometrically regular shape of the jets and desirable properties that derive from it.
Use of anti-$k_\rT$ jets persists in identifying jet flavour, which currently follow IRC unsafe flavour assignment procedures. 
As such, no robust comparison between data and the available precise fixed-order calculations can currently be carried out. 

The issue of IRC safety in the flavour assignment was first pointed out in Ref.~\cite{Banfi:2006hf,Banfi:2007gu} with a solution that modifies the jet definition itself to ensure IRC safety.
This algorithm, however, requires the flavour information of all particles as input, thus making an experimental realisation challenging.
Very recently, further approaches were proposed to assign heavy-flavour quantum numbers to jets: based on Soft Drop grooming techniques~\cite{Caletti:2022hnc}, through the alignment of flavoured particles along the Winner-Take-All axis~\cite{Caletti:2022glq}, or by modifying the anti-$k_\rT$ algorithm~\cite{Czakon:2022wam}.
Other prescriptions have also been proposed~\cite{Buckley:2015gua,Goncalves:2015prv,Ilten:2017rbd,Caletti:2021oor,Fedkevych:2022mid}.
However, no approach exists that both reproduces the same jets as a flavour-agnostic anti-$k_\rT$ algorithm, can be applied to generic processes with multiple jets, and at the same time is IRC safe to all orders.

In this work, we propose a new approach which allows to assign heavy-flavour quantum numbers to a set of flavour agnostic jets. This algorithm has the following properties:
(i)~it is IRC safe to all orders in perturbation theory and can therefore be applied in fixed-order predictions;
(ii)~it can be combined with any IRC safe definition of a jet, such as anti-$k_\rT$ jets, as the flavour assignment procedure is factorised from the jet reconstruction;
(iii)~the flavour assignment can be applied at the level of quarks, heavy-flavour hadrons, or with proxy particles that can be reconstructed in an experimental environment (such as secondary vertices, SVs).
The procedure we propose can therefore be directly applied in an experimental or theoretical setting to arbitrary scattering processes, enabling more direct and precise comparisons between theory and data.

In the following we present the algorithm, perform a test of IRC safety in a $\Pep\Pem$ environment, and focus on $\Pp\Pp\to\PZ+\Pqb\text{-}\jet$ production as a case study.

\sect{Inputs to the flavour dressing}
The proposed algorithm provides a way of assigning a specific flavour quantum number $\Fl$ to jets.
The required inputs to the algorithm are: a set of $m$ flavour-agnostic jets which have been obtained from an IRC safe jet definition, denoted by $\{j_1,\ldots,j_m\}$; 
the set of particles $\{p_1,\ldots,p_n\}$ and their flavour quantum numbers;
a criterion for associating particles with jets; and a flavour accumulation (counting) criterion.
Before presenting the algorithm, we first discuss these various inputs and comment on their potential differences in a theoretical or experimental setting.
\begin{itemize}[leftmargin=*]
  \item \textit{Flavour agnostic jets.} This set of inputs should be obtained with an IRC safe jet algorithm, and depending on the association criterion (see below) could require injecting ghost particles or retaining the constituent information of the jets.
  In the following, this set is considered to be composed of resolved ``analysis'' jets (either in exclusive or inclusive modes), that have passed a fiducial selection criterion.
  \item \textit{Particles and flavour.}
  In the first step, the set of particles can be identified with the same input that enters the flavour-agnostic jet algorithm.
  In a parton-level prediction, the subset of flavoured particles $f$ are identified as all (anti-)quarks in a given event with the flavour quantum number $\Fl$, e.g.\ $\Fl = \Pqc(\Pqb)$ when identifying $\Pqc(\Pqb)$-tagged jets.
  In a hadron-level prediction with stable heavy-flavour hadrons, the replacement $\Pqc(\Pqb)\to \PD(\PB)$ can be made.
  In an experimental setting, the flavoured particles can be replaced by a proxy particle for the heavy flavour, such as a reconstructed SV. In the latter, the charge information of the flavoured particles is likely unavailable and they may not have a definite flavour (i.e.\ they could have a probability of being associated to $\Pqc$ and $\Pqb$ heavy-flavours).
  \item \textit{Association criterion.} For each particle $p_i$ in the event, determine whether it can be associated to a jet. This criterion is important (although not unique) as only those jets that have at least one associated flavoured particle $f$ can be assigned non-zero flavour.
  From the point of view of a parton-level prediction, an obvious choice is to associate $p_i$ with $j_k$ if the corresponding particle $p_i$ is a constituent of a jet $j_k$.
  In cases where the flavoured particle $f_i$ does not directly enter the jet algorithm as input, other sensible options to make this association are: the requirement $\Delta R(f_i,j_k) < R_\tag$;
  or to include $f_i$ as ghost particles in the reconstruction to determine in which jets they are clustered with~\cite{Cacciari:2005hq,Cacciari:2008gn,ATL-PHYS-PUB-2015-013}.
  A discussion on the experimental feasibility of these approaches
  is given towards the end of this Letter.
  We emphasise that a jet flavour assignment based solely on this association criterion is \emph{not} IRC safe.
  \item \textit{Accumulation criterion.} In an ideal situation, both the flavour ($\Fl$) and charge ($\fl$ vs $\afl$) information of flavoured particles is known. In such a scenario, one considers $\fl(\afl)$ to carry a positive(negative) flavour quantum number, and an object is then considered flavoured if it is assigned an unequal number of $\fl$ and $\afl$.
  If the charge information is not available, one possibility is to instead consider an object to be flavoured if it has been assigned an odd number of flavoured particles $f$.
\end{itemize}

\sect{The flavour dressing algorithm}
With this information at hand, the flavour dressing algorithm to identify whether a reconstructed jet can be assigned the flavour quantum number $\Fl$ proceeds as follows:
\begin{enumerate} [leftmargin=*]
  \item Initialise empty sets $\tag_k = \varnothing$ for each jet $j_k$ to accumulate all flavoured particles assigned to it.
  \item Populate a set \dset of distance measures based on all allowed pairings:
  \begin{enumerate}
    \item For each unordered pair of particles $p_i$ and $p_j$, add the distance measure $d_{p_ip_j}$ if either
    \emph{both particles are flavoured}%
    \footnote{If the charge information is available, the pairings can be restricted to only compatible quantum numbers, i.e.\ $(\fl,\afl)$ but not $(\fl,\fl)$.}
    or \emph{at least one particle is un-flavoured and $p_i$ and $p_j$ are associated with the same jet}.
    \item If the particle $p_i$ is associated to jet $j_k$, add the distance measure $d_{p_i j_k}$.
    In a hadron collider environment, the beam distances $d_{p_i B_{\pm}}$ should be added if $p_i$ is not associated to any jet.
  \end{enumerate}
  \item While the set \dset is non-empty, select the pairing with the smallest distance measure:
  \begin{enumerate}
    \item $d_{p_i p_j}$ is the smallest:
    the two particles merge into a new particle $k_{ij}$ carrying the sum of the four-momenta and flavour. All entries in \dset that involve $p_i$ or $p_j$ are removed and new distances for $k_{ij}$ are added.
    \item \(d_{p_i j_k}\) is the smallest: assign the particle $p_i$ to the jet $j_k$, $\tag_k \to \tag_k \cup \{p_i\}$, and remove all entries in \dset that involve $p_i$.
    \item \(d_{p_i B_{\pm}}\) is the smallest: discard particle $p_i$ and remove all entries in \dset that involve $p_i$.
  \end{enumerate}
  \item The flavour assignment for jet $j_k$ is determined according to the accumulated flavours in $\tag_k$.
\end{enumerate}
For the distance measure between two final-state objects $a$ and $b$ (particles or jets) we use
\begin{subequations}
\label{eq:dist}
\begin{align}
  \label{eq:dist_ab}
  d_{ab} &= \Omega_{ab}^2 \;
  \max\bigl(p_{\rT,a}^\alpha,p_{\rT,b}^\alpha\bigr) \;
  \min\bigl(p_{\rT,a}^{2-\alpha},p_{\rT,b}^{2-\alpha}\bigr) \, ,
  \\
  \intertext{with}
  \Omega_{ab}^2 &= 2\,\biggr[\frac{1}{\omega^2}(\cosh(\omega \Delta y_{ab}) - 1) - (\cos \Delta \phi - 1)\biggr]\,. \nonumber
\end{align}
If not stated otherwise, the choice of values for the parameters are: $\alpha=1$ and $\omega=2$.
For lepton colliders, this measure can be adjusted through a suitable replacement of variables~\cite{Catani:1993hr,Ellis:1993tq}.
The distance between particle $p_i$ and a hadron beam in the direction of positive $(+)$ or negative $(-)$ rapidity is
\begin{align}
  \label{eq:dist_fB}
  d_{p_i B_{\pm}} =
  \max\bigl(p_{\rT,i}^\alpha,p_{\rT,B_{\pm}}^\alpha(y_{i}) \bigr) \;
  \min\bigl(p_{\rT,i}^{2-\alpha},p_{\rT,B_{\pm}}^{2-\alpha}(y_{i})\bigr) \,,
  \nonumber\\
  p_{\rT,B_{\pm}}(y) =
  \sum_{k=1}^m p_{\rT,{j_k}} \Bigl[
  \Theta( \pm \Delta y_{j_k}) +
  \Theta( \mp \Delta y_{j_k}) \; \re^{\pm \Delta y_{j_k}}
  \Bigr] \,,
\end{align}
\end{subequations}
with the rapidity separation $\Delta y_{j_k} = y_{j_k} - y$ and $\Theta(0) = \tfrac{1}{2}$.
The distance measures in Eq.~\eqref{eq:dist} are inspired by the flavour-$k_\rT$ algorithm~\cite{Banfi:2006hf} and its generalisation in \cite{Caola:2023wpj}.
This choice of measure ensures that soft pairs of flavoured particles are recombined early on, thus avoiding a sensitivity to infrared physics.
A hierarchical tagging of flavours can also be applied, e.g.\ by running the algorithm for $\Fl = \Pqb$ and then $\Fl = \Pqc$ and requiring that $\Pqc$-jets must not have a $\Pqb$-flavour assignment.
In principle, although not considered in the following, the algorithm could also be operated when considering several flavours such as $\Fl = \Pqc, \Pqb$ with an appropriate adjustment to the accumulation criterion.

\sect{Test of IRC safety in $\Pep\Pem\to\jets$}
In order to test the IRC safety of the flavour-dressing algorithm, a resolution variable is introduced that allows to probe the fully unresolved regimes, i.e.\ restricting all emissions to be only soft and/or collinear.
In this limit, the probability of a mis-identification of flavours (a ``bad'' tag) must vanish for any IRC-safe procedure of identifying jet flavour.

\begin{figure}
  \centering
  \includegraphics[width=.75\linewidth, trim=2 0 9 0, clip]{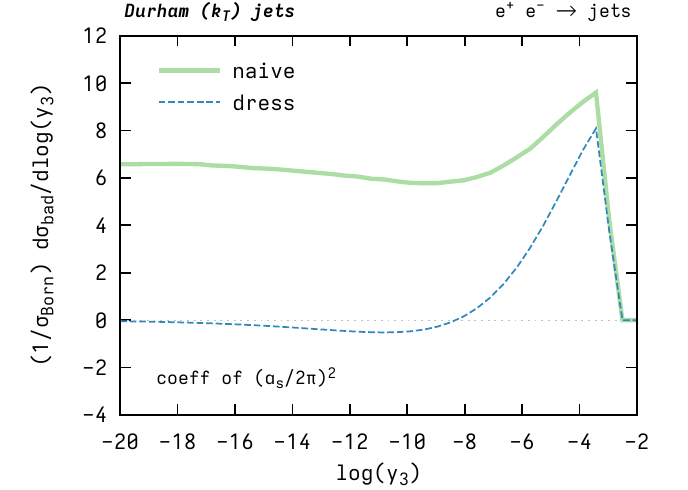}%
  \\
  \includegraphics[width=.75\linewidth, trim=2 0 9 0, clip]{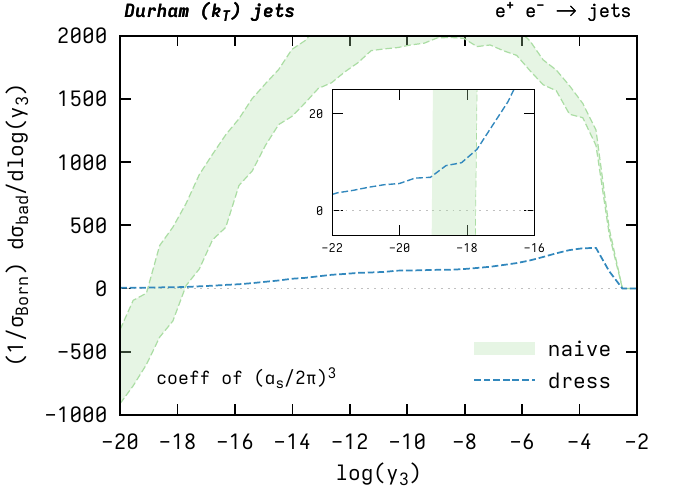}%
  \caption{\label{fig:epem_ir_test}%
    Behaviour of the flavour mis-identified (``bad'') cross section in $\Pep\Pem\to\jets$ production as a function of the $y_3$ resolution variable.
    Comparison of a naive flavour assignment~(green) with the flavour-dressing approach~(blue, orange) at the $2^\text{nd}$~(upper) and $3^\text{rd}$~(lower) order in $\alphas$.
  }
\end{figure}

For the $\Pep\Pem\to\jets$ process, the correct flavour assignment in the unresolved limit is determined by the underlying Born-level scattering reaction, $\Pep\Pem \to \fl \afl$, and therefore corresponds to two jets with a net flavour-tag.
Jets are defined using the $k_\rT$ (or ``Durham'') algorithm~\cite{Catani:1991hj}, which is not IRC safe in the case of a naive flavour assignment, i.e.\ simply accumulating the flavours of the jet constituents.
A suitable resolution variable for this process is given by the parameter $y_3$, which determines the transition between identifying an event as a 2-jet or a 3-jet configuration in the Durham algorithm.
As such, it allows to probe the fully unresolved region by inspecting the limit $y_3 \to 0$.

In Fig.~\ref{fig:epem_ir_test} we perform a comparison between different prescriptions of assigning flavour to the jets as a function of the $y_3$ resolution variable.
For simplicity, the test is performed by considering all (anti-)quarks to carry a single quantum number $\fl(\afl)$.
These comparisons are provided for the perturbative coefficients of the cross-section up to third order, i.e.\ $\mathcal{O}(\alphas^3)$, using the calculation of Refs.~\cite{Gehrmann-DeRidder:2014hxk,Gehrmann:2017xfb}.
At first order (not shown), the $\Pep\Pem \to \fl \afl \Pg$ process is not yet exposed to the subtleties of flavour creation that jeopardises IRC safety and also the naive prescription is thus IRC safe.
Starting from the second order, however, the naive prescription develops a soft singularity,
which manifests itself by the associated curve (solid green) in the upper figure approaching a non-vanishing value in the $y_3\to0$ limit.
At third order, the IRC un-safe behaviour of the naive prescription becomes more severe as can be seen in the lower plot; the IRC singularities in this case are no longer confined to the $y_3\to0$ regime but the entire spectrum is ill defined as indicated by the width of the green band that corresponds to varying the internal technical cut-off parameter of the calculation.
The flavour dressing approach (solid blue and dashed orange), on the other hand, correctly approaches zero in the limit $y_3 \to 0$ at all considered orders, confirming the IRC safety of the procedure.

\sect{Application to $\PZ+\Pqb\text{-}\jet$ production}
Beyond the IRC safety test discussed so far, it is also important to apply and test the flavour dressing algorithm in realistic scenarios.
To do so, we consider the process $\Pp\Pp \to \PZ+\Pqb\text{-}\jet$, and compare theory predictions based on fixed order (parton-level) with those obtained by matching fixed-order predictions with a Parton Shower (PS) Monte Carlo. Comparisons of these predictions, for a range of differential observables, demonstrate the potential sensitivity of the algorithm to universal all-order effects and non-perturbative corrections.

The fixed-order parton-level predictions are obtained up to \NNLO~\cite{Gauld:2020deh}, and are compared to hadron-level \NLOPS accuracy generated with \aMCatNLO~\cite{Alwall:2014hca} interfaced to \pythia\,8.3~\cite{Bierlich:2022pfr}.
The NNPDF3.1 \NNLO PDF set~\cite{NNPDF:2017mvq}, accessed via LHAPDF~\cite{Buckley:2014ana}, with $\alpha_s(M_\PZ) = 0.118$ and $n_f^{\rm max} = 5$ is used throughout. 
The complex-mass and $G_{\mu}$ input schemes are adopted using the values as quoted in~\cite{ParticleDataGroup:2018ovx}.
The central prediction is obtained for the central scale $\mu_0 \equiv E_{\rT,\PZ}$, and uncertainties due to the variation of the factorisation ($\mu_F$) and renormalisation ($\mu_R$) scales by a factor of two around $\mu_0$, with the constraint $\frac{1}{2} \leq \mu_F/\mu_R \leq 2$, are shown as bands. 
The following fiducial selection is applied:
$m_{\ell\bar\ell} \in [71,111]\,\GeV$, $p_{\rT,\ell(j)} > 27(40)\,\GeV$, $|\eta_{\ell/j}| < 2.5$, and $\Delta R(\ell,j) > 0.4$.
The set of $R=0.4$ anti-$k_\rT$ jets passing this fiducial selection are then used as an input to the flavour-dressing algorithm described in this Letter.
In the final selection, we additionally require the leading jet to be flavoured.

\begin{figure*}
  \centering
  \hfill
  \includegraphics[width=.33\linewidth, trim=2 0 9 0, clip]{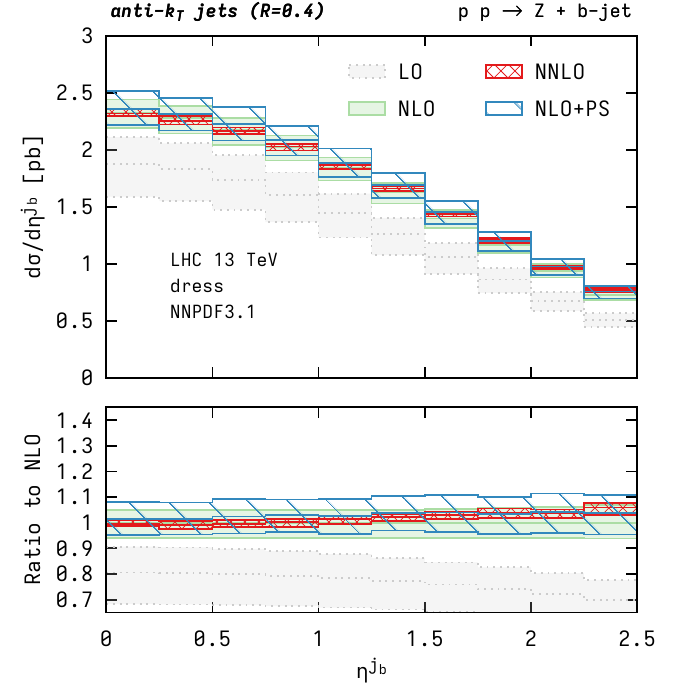}%
  \hfill
  \includegraphics[width=.33\linewidth, trim=2 0 9 0, clip]{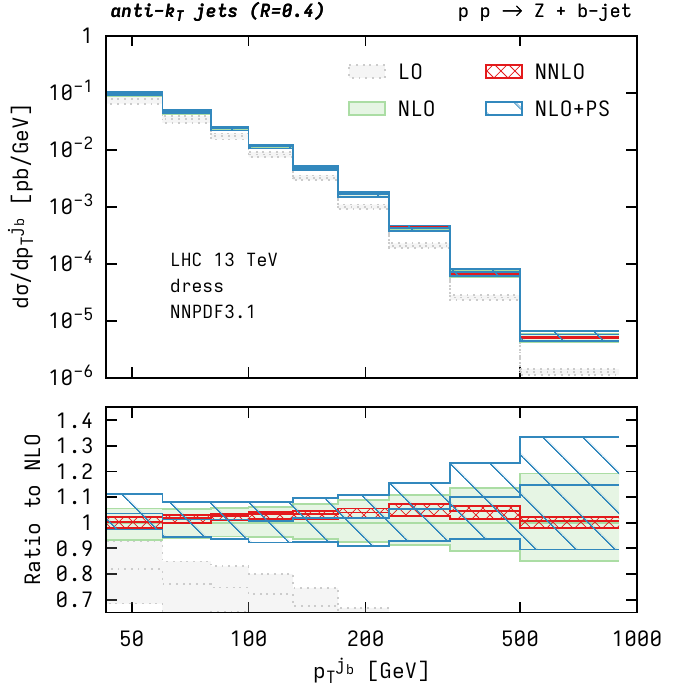}%
  \hfill
  \includegraphics[width=.33\linewidth, trim=2 0 9 0, clip]{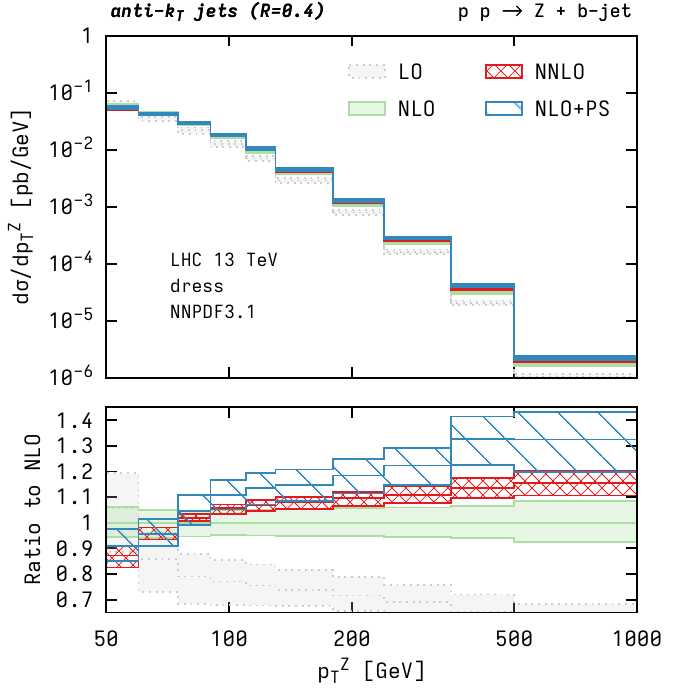}%
  \hfill\smash{ }
  \caption{Comparison between fixed-order and \NLOPS descriptions of the leading flavoured jet pseudorapidity (left), transverse momentum (central), and the $\PZ$ boson transverse momentum (right). Upper panels show the absolute cross-section distributions, while those in the lower panels are normalised to the central \NLO prediction. Scale uncertainties are shown in each case.}
  \label{fig:zb}
\end{figure*}

The results of the comparison are shown in Fig.~\ref{fig:zb} for the pseudorapidity ($\eta_\Pqb$) and transverse momentum of the leading flavoured jet ($p_{\rT,\Pqb}$) and $\PZ$ boson ($p_{\rT,\PZ}$).
A good agreement between the fixed-order and hadron-level predictions is found, both in terms of overall normalisation and shape of the distributions.
At \NLO accuracy the agreement is typically within 2\%, demonstrating that the algorithm is robust with respect to both non-perturbative (hadronisation corrections) and universal all-order effects. 
The latter is again verified by the fact that the \NNLO accurate distribution lies within the uncertainty estimate of both the \NLO and \NLOPS predictions.
In the case of $p_{\rT,\PZ}$ some sensitivity to higher-order corrections is observed, which is indicated by the poorer agreement between \NLO and \NLOPS predictions (although typically compatible within uncertainties).
This effect can be traced back to flavour-creation through gluon emission with a subsequent $\Pg\to\fl\afl$ splitting, which cannot yet be accessed in a fixed-order \NLO prediction. Indeed, we observe that at \NNLO the description is greatly improved.
As a further test of the IRC safety of the algorithm, it was checked that the \NNLO calculation was independent of the internal technical cut-off parameter.

\sect{Experimental feasibility at the LHC}
As demonstrated above, the flavour dressing procedure is IRC safe, generally applicable to IRC safe flavour-agnostic jets, and may enable more precise and direct theory--data comparisons in the future.
Such comparisons will rely on the proposed algorithm being successfully applied to data, a task which must be carried out by experimental collaborations.
In the following we provide several comments on the feasibility of this.

The proposed flavour dressing algorithm is applicable to the anti-$k_\rT$ jets which are already used by LHC collaborations, and are already the input to current (IRC unsafe) flavour assignment procedures. Therefore, no modification of the actual jet reconstruction is required to apply the flavour dressing approach.

The association criteria of the flavour dressing algorithm are also applicable to secondary vertices (SV).
If the SV objects are not inputs to the initial jet reconstruction---as is a common experimental approach at the LHC~\cite{LHCb:2015tna,ATLAS:2015thz,CMS:2017wtu,LHCb:2021dlw}---one can still determine whether the decay products of these objects are constituents of the jets to perform an association. In ambiguous cases (e.g. the decay products of one SV appear as constituents of more than one jet), a unique assignment can still be made after introducing a measure between the SV and jets.
Such a measure could also be used for the $\Delta R(f_i,j_k) < R_\tag$ association criterion, in cases where this condition was satisfied by multiple jets.
Alternatively, the association could be made by incorporating the information of the SV clusters through the use of ghost particles.

The SV objects may carry a probability of being associated to a bottom, charm, or light flavours.
In this case, the flavour dressing algorithm could be applied by probabilistically assigning a specific and single flavour to the SV.
Given that the number of reconstructed SVs for a given event is small (for the $\PZ+\Pqb\text{-}\jet$ scenario considered above, the mean value of SVs per event was close to two), it may also be possible to propagate the flavour information of the SVs in a systematic way.

Ideally, the flavour dressing algorithm captures all flavoured particles in an event. Due to the finite fiducial volume of the detector, and that not all heavy-flavour hadrons decay modes will generate a detectable SV, this will not be possible for all events.
An experimental efficiency correction will be required to account for this and for the possibility of ``fakes'', which is standard for the experimental collaborations.

Finally, we note that the proposed algorithm opens up the possibility for a number of experimental (and theoretical) studies related to the properties/structure of flavoured jets, which is left for future work.

\sect{Conclusions and Outlook}
In this Letter, a novel approach is proposed for assigning (heavy-)flavour quantum numbers to arbitrary flavour-agnostic jets that is IRC safe to all orders.
The \emph{flavour dressing} algorithm accomplishes this by fully disentangling the kinematic reconstruction of jets from the flavour assignment, giving rise to a simple yet very generic approach that can be applied universally to the study of all physics processes that involve flavoured jets.
While specific choices were made both in the association and accumulation criterion, alternatives (some of which we noted) can be considered in view of experimental feasibility.

The property of IRC safety is imperative for a robust theoretical definition of flavoured jet observables and was explicitly tested for $\Pep\Pem$ and $\Pp\Pp$ environments---up to order $\mathcal{O}(\alphas^3)$. The sensitivity of the algorithm to universal all-order effects was also tested in a $\Pp\Pp$ environment.
While the issues of IRC safety are deeply linked to the use of massless quarks in the calculation, such a setup is the basis for the resummation of potentially large mass logarithms $\ln(Q^2/m_\Pq^2)$ to all orders that are otherwise only accounted for to a finite order when quark masses $m_\Pq$ are retained in the calculation. 
Moreover, an IRC unsafe prescription introduces a direct sensitivity to such mass logarithms that are not power suppressed and can thus potentially spoil the reliability of the calculation.

As an explicit example, the flavour dressing algorithm was applied to the process $\Pp\Pp \to \PZ+\Pqb\text{-}\jet$ highlighting its robustness with respect to non-perturbative hadronisation effects and higher-order corrections as modelled by parton showers. 
Allowing for a theoretically rigorous flavour-tagging of anti-$k_\rT$ jets, the flavour dressing algorithm further resolves the main mismatch between theory and data, paving the way for precision phenomenology using flavoured jets.

\sect{Acknowledgements}
We thank Aude Gehrmann--De Ridder, Thomas Gehrmann and Nigel Glover for their input and encouragement to pursue this work.
We are grateful to Fabrizio Caola, Radosław Grabarczyk, Max Hutt, Gavin Salam, Ludovic Scyboz, and Jesse Thaler for a critical assessment of our algorithm, in particular exchanges on flavoured object inputs; Tancredi Carli and Joey Huston for discussions on the experimental feasibility; Phil Ilten for comments and advice/details on the use of SV objects in an experimental environment; Andrea Banfi, Andrew Larkoski, Simone Marzani, Pier Monni, Daniel Reichelt, and Giulia Zanderighi for discussions and/or comments on the manuscript.
This work was supported in part by the Swiss National Science Foundation (SNF) under contract 200020-204200.

\sect{Note added}
In this version of the manuscript, we have corrected issues in the original flavour-dressing algorithm that led to an IRC unsafety as was pointed out~\cite{Caola:2023wpj}.
We thank the authors of \cite{Caola:2023wpj} for communication and providing their IRC testing suite to check our revised algorithm.

\bibliography{jetflav}

\end{document}